\documentclass[twocolumn,showpacs,preprintnumbers,amsmath,amssymb,floatfix,superscriptaddress]{iopart}
\usepackage{amsfonts}
\usepackage{amssymb}
\usepackage{graphicx}
\usepackage{dcolumn}
\usepackage{bm}
\usepackage[dvips]{color}
\usepackage[displaymath]{preview}

\begin{document}

\title{An eccentrically perturbed Tonks-Girardeau gas}

\author{J.~Goold}
%\email{cqtjg@nus.edu.sg}
\address{Centre for Quantum Technologies, National University of Singapore,
             Singapore, Singapore}
\author{M.~Krych, Z.~Idziaszek }
\address{Faculty of Physics, University of Warsaw, 00-681 Warsaw, Poland}
\author{T.~Fogarty, Th.~Busch}
\address{Department of Physics, University College Cork, Cork, Ireland}

\date{\today}

\begin{abstract}
  We investigate the static and dynamic properties of a
  Tonks-Girardeau gas in a harmonic trap with an eccentric
  $\delta$-perturbation of variable strength. For this we first find
  the analytic eigensolution of the single particle problem and use
  this solution to calculate the spatial density and energy profiles
  of the many particle gas as a function of the strength and position
  of the perturbation. We find that the crystal nature of the Tonks
  state is reflected in both the lowest occupation number and momentum
  distribution of the gas.  As a novel application of our model, we
  study the time evolution of the the spatial density after a sudden
  removal of the perturbation. The dynamics exhibits collapses and
  revivals of the original density distribution which occur in units
  of the trap frequency. This is reminiscent of the Talbot effect from
  classical optics.
\end{abstract}

\maketitle

\section{Introduction}
% General introduction finishing with one dimensional gases
One of the standout achievements of ultracold atomic physics in the
past decade has been the realisation of controllable and clean
environments for the simulation of quantum manybody systems
\cite{Bloch:08}. Cold atoms are ideal systems for such simulations and
offer many desirable features, two of which are particularly
unique. The first of these is the ability to tune the scattering
length and hence the interaction in quantum gases. This is achieved by
means of Feshbach resonances \cite{Courteille:98}. The second is the
ability to change the dimensionality of the system and in particular
to generate periodic geometries for atoms by means of optical lattice
potentials \cite{Greiner:02}. Both of these separate developments, or
their combination, allows experimentalists to enter regimes of strong
correlation and explore complex condensed matter models. For instance,
using optical lattice potentials the superfluid-Mott insulator phase
transition has been observed by changing the lattice depth
\cite{Greiner:02,Jaksch:98}. In addition, by loading a BEC into a
two-dimensional lattice and utilising strong confinement in the
transverse direction, the achievement of lower dimensional systems was
made possible and new quantum phases can now be explored with an
incredible degree of precision. In the case of two dimensions, the
famous Berezinskii-Kosterlitz-Thouless phase transition was observed
recently \cite{Hadzibabic:06} and for ultracold gases in one dimension
one of the crowning achievements was the realisation of a unique model
from many body theory \cite{Girardeau:60} - the strongly correlated
Tonks-Girardeau gas of hard core bosons \cite{Parades:04,Weiss:04}.

Since the first creation of a Tonks-Girardeau gas there have been
further experimental studies focusing on the absences of
thermalisation due to the integrability of the underlying Hamiltonian
\cite{Kinoshita:06} and more recently on the density dynamics in the
presence of an outcoupled impurity \cite{Kohl:09}. This second work
has opened the door to studying the transport properties of impurities
in ultracold quantum gases and therefore highlights the importance to
understand the fundamental properties of a Tonks-Girardeau gas in the
presence of an impurity. A number of studies in this direction already
exist in the literature, in which the impurity is modelled using
pseudo-potential approximation
\cite{Busch:03,Fu:06,Goold1:08,Goold2:08,Girardeau:09,Goold:10}. These
studies have primarily focused on fixed position perturbations and in
this work we introduce a versatile analytical model which can be used
to describe the Tonks-Girardeau gas in the presence of a perturbation
of arbitrary strength at any position in a harmonic trap. In addition
to describing a static impurity, this model can be interpreted as the
limiting case of an split, asymmetric double well trap which may be
realized using a sharply focused laser beam which is detuned from the
atomic transition.

The paper is organised as follows: in Sec.~\ref{sect:tggas} we briefly
remind the reader of the Tonks-Girardeau gas and its solution by the
Fermi-Bose mapping theorem. We then outline the eigensolution of a
harmonically trapped single particle in the presence of an off-centre
$\delta$-perturbation in Sec.~\ref{sect:sp} and show in
Sec.~\ref{sect:properties} how this can be used in combination with
known techniques to characterise the groundstate of a many-particle
gas. We then study the dynamics of the single particle density after a
sudden removal of the perturbation as well as for a time-of-flight
measurement. Finally, we conclude.

\section{The Tonks-Girardeau gas}
\label{sect:tggas}

Optical lattices and atom chips allow to create trapping potentials
that are tight enough in the transversal direction to freeze out all
dynamics in these degrees of freedom \cite{Moritz:03}. A gas of $N$
bosons in such a potential can then be approximated by the
one-dimensional Hamiltonian
\begin{eqnarray}
 \label{eq:TG_ham}
 \mathcal{H}=\sum_{n=1}^N\left[
      -\frac{\hbar^2}{2m}\frac{\partial^2}{\partial x_n^2}
      +V_{ext}(x_{n})\right]
      +g_{1D}\sum_{i<j}\delta(|x_i-x_j|)\;,
\end{eqnarray}
where $m$ is the mass of the particles, $V_{ext}$ is the trapping
potential and $g_{1D}$ is a 1D coupling constant which is derived from
the renormalisation of the three-dimensional scattering process
$g_{1D}=\frac{4\hbar^2 a_{3D}}{ma_\perp} \left(a_\perp-Ca_{3D}
\right)^{-1}$ \cite{Olshanii:98}. Here $a_\perp$ is the trap width and
$C$ is a constant of value $1.4603...$. This Hamiltonian describes an
inhomogeneous Lieb-Liniger gas, which in the strongly repulsive limit,
$g_{1D}\rightarrow\infty$, can be solved by using a mapping to an
ideal and spinless fermionic system \cite{Girardeau:60}. This
procedure is known as the Fermi-Bose mapping theorem and it can be
used to show that the \textit{local} density and correlation functions
of this \textit{strongly correlated} system are equivalent to the
corresponding quantities of a \textit{ non-interacting} spin polarized
Fermi gas. This strange equivalence is due solely to the
dimensionality of the system - as the repulsive interactions become
stronger, the particles are no longer free to overlap, thus mimicking
the Pauli-exclusion principle in configuration space. The essential
idea is that one can then treat the interaction term in
eq.~(\ref{eq:TG_ham}) by replacing it with a boundary condition on the
allowed bosonic wave-function
\begin{equation}
  \label{eq:constraint}
  \Psi_B(x_1,x_2,\dots,x_n)=0\quad \mbox{if} \quad |x_i-x_j|=0\;,
\end{equation}
for $i\neq j$ and $1\leq i\leq\ j\leq N$.  This is simply the hard
core constraint which says that no probability exists for two
particles ever to be at the same point in space. Such a constraint is
automatically fullfilled by calculating the wave-function using a
Slater determinant
\begin{equation}
  \Psi_F(x_1,x_2,\dots,x_N) =\frac{1}{\sqrt N!}
                             \det_{(n,j)=(0,1)}^{(N-1,N)}\psi_n(x_j)\;,
\label{eq:psiF}
\end{equation}
where the $\psi_n$ are the single particle eigenstates of the ideal
system. This, however, leads to a fermionic rather than bosonic
symmetry, which can be corrected by a multiplication with the
appropriate unit antisymmetric function \cite{Girardeau:60}
\begin{equation}
  A=\prod_{1\leq i < j\leq N} \mbox{sgn}(x_i-x_j)\;,\\
  \label{eq:mapFB}
\end{equation}
to give $\Psi_B=A \Psi_F$. Knowing the single particle eigenstates is
therefore a pre-requisite to being able to apply the mapping
mechanism.

\section{Single particle problem}
\label{sect:sp}
The number of models in many particle quantum physics for which
analytical solutions exist is extremely limited and often restricted
to one spatial dimension \cite{Sutherland}. Here we will describe one
such model by studying the Hamiltonian (\ref{eq:TG_ham}) with
\begin{equation}
  \label{eq:extV}
  V_{ext}(x)=\frac{1}{2}m\omega^2 x^2 + \gamma\delta(x-d)\;,
\end{equation}
and $g_{1D}\rightarrow\infty$, which describes a Tonks-Girardeau gas
in a harmonic trap of frequency $\omega$ which is split asymmetrically
by a tunable $\delta$-perturbation of strength $\gamma$ a distance $d$
from the centre of the trap. The limit $d=0$ for this model is well
known and has been studied extensively in recent years
\cite{Busch:03,Goold1:08,Goold2:08, Murphy:07}. In order to use the
prescription of the mapping theorem one needs to first solve for the
single particle eigenfunctions to build the Slater determinant
(\ref{eq:psiF}). In this section we will study the eigensolution of
the single particle problem in detail. In fact, this problem is
equivalent to the relative problem of two interacting atoms in
seperate harmonic traps which has recently been numerically solved in
the three dimensional case \cite{Krych:09}. We show below that this
problem is analytical in one dimension.

The single particle Schr\"{o}dinger equation for the potential
(\ref{eq:extV}) is given by
\begin{equation}
  \label{eq:sp}
  \left(\frac{-\hbar^2}{2m}\frac{d^2}{dx^2}
        +\frac{1}{2}m\omega^2 x^2 
        + \gamma\delta(x-d)\right) \psi_{\nu}(x)=\epsilon_{\nu}\psi_{\nu}(x)\;,
\end{equation}
    where the energies are given by
    $\epsilon_{\nu}=(\nu+\frac{1}{2})\hbar\omega$. Let us rescale all
    of the units in terms of the undisturbed ($\gamma=0$) trap length
    $l_{0}=\sqrt{\frac{\hbar}{m\omega}}$ and energy $\hbar\omega$. In
    this way the equation (\ref{eq:sp}) can be rewritten as
\begin{equation}
  \label{eq:de}
  \left(\frac{d^2}{dx^2}+\nu +\frac{1}{2}-\frac{x^2}{2}-\kappa\delta(x-d)\right)
  \psi_{\nu}(x)=0,
\end{equation}
where $\kappa = \gamma l_0/(\hbar \omega)$ is the
re-normalised strength of the $\delta$-barrier.  On either side of the
$\delta$-function the solutions of the differential equation
(\ref{eq:de}) are parabolic cylinder functions $D_{\nu}(x)$, which
vanish for $x \rightarrow \infty$, but diverge for $x \rightarrow
-\infty$.  We can therefore write the solution piecewise as
\begin{equation}
\label{eq:spstates}
\psi_{\nu}(x)=\psi_{l,\nu}(x)\theta(d-x)+\psi_{r,\nu}(x)\theta(x-d)\;,
\end{equation}
with
\begin{equation}
   \psi_{r,\nu}(x)=N_+D_\nu( x)\quad \mbox{and} \quad 
   \psi_{l,\nu}(x)=N_-D_\nu(-x)\;.
\end{equation}
and $\theta(x)$ being the Heaviside function. The condition of
continuity of these solutions at the position of the $\delta$-function
\begin{equation}
  \label{eq:cont}
  N_+D_\nu(-d)=N_-D_\nu(d)\;,
\end{equation}
together with the solution of the Schr\"odinger equation
\begin{equation}
    -\int_{d-\epsilon}^{d+\epsilon} \psi_\nu^{''}(x) dx
    +\int_{d-\epsilon}^{d+\epsilon} V(x)\psi_\nu(x) dx 
 = \epsilon_\nu \int_{d-\epsilon}^{d+\epsilon} \psi_\nu(x) dx\;,
\end{equation}
with
\begin{equation}
  V(x)=\frac{1}{2} x^2+\kappa\delta(x-d)\;,
\end{equation}
leads to a transcendental equation which determines the energy
eigenvalues as a function of both, $\kappa$ and $d$
\begin{equation}
  N_{+}D'_{\nu}(-d)+N_{-}D'_{\nu}(d) -2 \kappa N_{+}D_\nu(-d)=0\;.
  \label{eq:conn}
\end{equation}
The derivatives of the parabolic cylinder functions can be calculated
using the recurrence relation \cite{Abramowitz:72}
\begin{equation}
  D'_\nu (z)=\nu D_{\nu-1} (z) -\frac12 z D_\nu (z)\;,
  \label{eq:dif}
\end{equation}
and for $D_\nu(d) \neq 0$ we find from eq.~(\ref{eq:cont}) the
relation $N_{-}=N_{+} D_{\nu}(-d) /D_{\nu}(d)$. Substituting these
into eq.~(\ref{eq:conn}) we therefore find
\begin{eqnarray}
  \nu \Big[ D_{\nu-1}(-d)D_\nu(d)+D_{\nu-1}&(d) D_{\nu}(-d)\Big]=2\kappa D_{\nu}(-d)D_{\nu}(d)\;.
\end{eqnarray}
This equation determines the eigenenergies for the solutions that are
nonzero at the position of the delta potential. To find the ones for
which $D_\nu(d)= 0$ we can see from eq.~(\ref{eq:cont}) that
$D_\nu(-d)= 0$, provided that $N_{+}\neq 0$ and $N_{-} \neq 0$. Then
eq.~(\ref{eq:conn}) reduces to
\begin{equation}
  \label{eq:sol2}
  N_{+} \nu D_{\nu-1}(-d)+N_{-} \nu D_{\nu-1}(d)=0.
\end{equation}
where the derivatives have been determined using
eq.~(\ref{eq:dif}). The above equation together with the normalization
of the total wave function determines $N_+$ and $N_-$ for the
solutions that vanish at the position of the delta potential. These
solutions are independent of $\kappa$, and they correspond to the
harmonic oscillator wave functions that vanish at
$x=d$. 
\begin{figure}[tb]
  \begin{center}
    \includegraphics[width=\linewidth , clip] {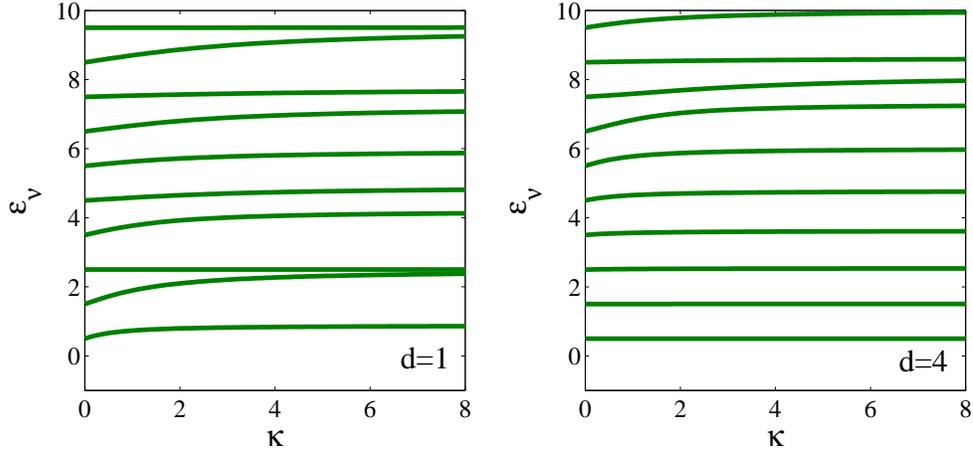}
  \end{center}
  \caption{The energy spectrum of harmonically trapped particle in the
    presence of $\delta$-like perturbation at position $d=1$ and $d=4$, for different strengths, $\kappa$, of the perturbation.}
  \label{fig:EvsD}
\end{figure}

\begin{figure}[tb]
  \begin{center}
    \includegraphics[width=\linewidth , clip] {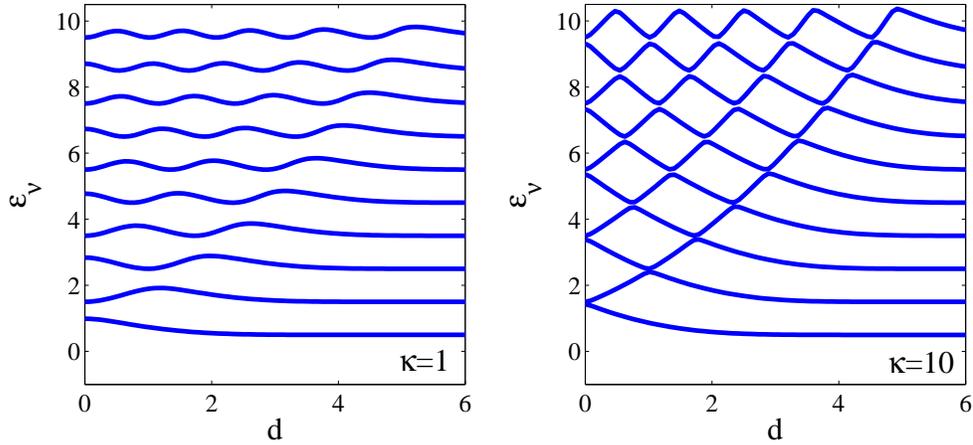}
  \end{center}
  \caption{The energy spectrum of harmonically trapped particle in the
    presence of $\delta$-like perturbation with strength $\kappa=1$ and $\kappa=10$,
    for various positions of the perturbation.}
  \label{fig:EvsK}
\end{figure}
\begin{figure}[tb]
  \begin{center}
    \includegraphics[width=\linewidth , clip] {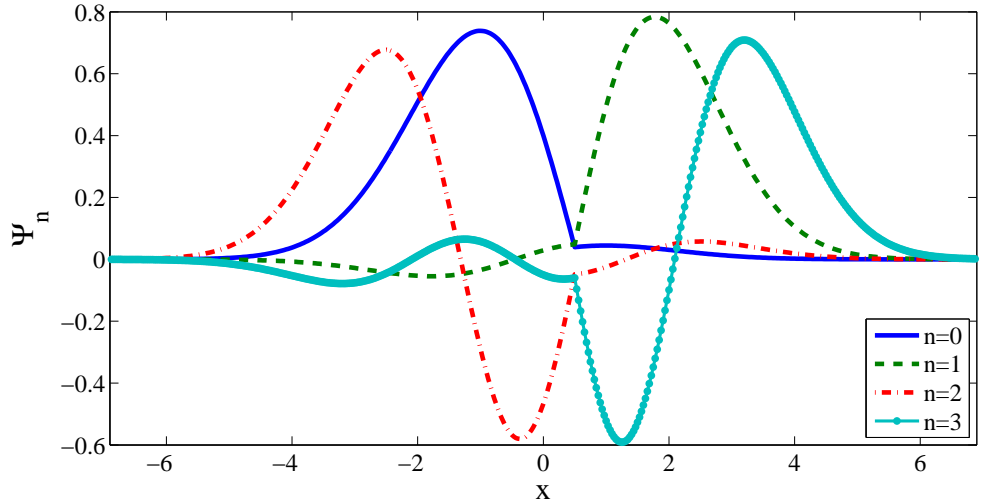}
  \end{center}
  \caption{The lowest four eigenfunctions of a particle in a harmonic
    potential with a repulsive $\delta$-function located at $d=0.5$ of
    strength $\kappa=10$}
  \label{fig:Wave}
\end{figure}

In Figs.~\ref{fig:EvsD} and \ref{fig:EvsK} we show the energy spectrum
as a function of $d$ and $\kappa$, respectively. The presence of the
perturbation introduces a non trivial structure in the spectrum and in
general leads to an increase of the state's energy. The four lowest
lying eigenfunction for $d=0.5$ and $\kappa=10$ are shown in
Fig.~\ref{fig:Wave}. It can be seen that the presence of delta
potential results in an abrupt change in the slope of the wavefunction
at the position of the delta function and that the states are already
highly localised on one side of the impurity for $\kappa=10$.

\section{Static and Dynamic Properties}
\label{sect:properties}

\subsection{Single Particle Density Profile}
\label{sect:spd}
The single particle density is one of the most important observables
for ultracold quantum gases. In the Tonks regime one can obtain it,
even time-dependently, from the spectrum of underlying single particle
Hamiltonian as \cite{Girardeau:00}
\begin{eqnarray}
  \label{eq:spd}
  \rho(x,t)&=N \int_{-\infty}^{+\infty}|\Psi_B (x,x_2,\dots,x_N;t)|^2dx_2\dots dx_N
             \nonumber\\
           &=\sum_{n=0}^{N-1}|\psi_{n}(x,t)|^2\;,
\end{eqnarray}
where we have adopted the convention of labeling the first $N$
eigenfunctions as $n=0,1,2,\dots N$.

\begin{figure}[tb]
  \begin{center}
    \includegraphics[width=\linewidth , clip] {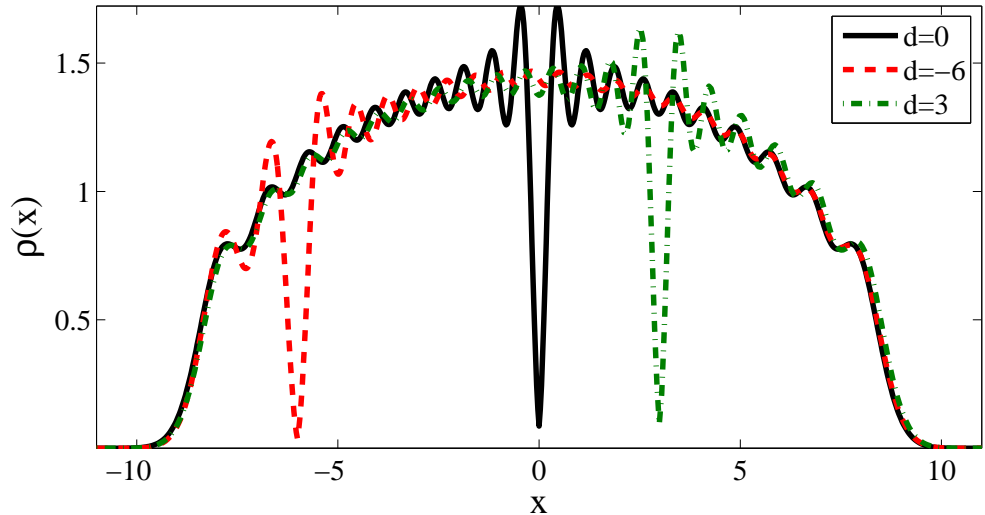}
  \end{center}
  \caption{ The single particle density of a harmonically trapped
    Tonks-Girardeau gas with 20 particles in equilibrium with a
    barrier of strength $\kappa=10$, located at positions $d=0,3,-6$.}
  \label{fig:spd}
\end{figure}

In Fig.~\ref{fig:spd} we show this single particle density for a gas
of 20 particles in a trap with an impurity of strength $\kappa=10$ at
three different positions in the trap. As expected the
$\delta$-impurity creates a significant local dip in the density and
has only minimal effect at larger distances. The enhanced oscillations
which are present around the position of the impurity are analogous to
the famous Friedel oscillations which occur around an impurity in the
surface charge density of a homogeneous electron gas
\cite{Friedel:58}.

Let us in the following investigate the dynamical properties of such a
gas by examining a non-equilibrium situation created by sudden removal
of the impurity. In order to compute the time-dependent density of
eq.~(\ref{eq:spd}) one needs to employ a time dependent basis. We
obtain this basis numerically using the split operator method in the
unperturbed harmonic trap. Alternatively one may simply employ the
well known propagator for the harmonic oscillator to get the time dependent basis \cite{Feynman}. This can then be used to calculate the single particle density of
eq.~(\ref{eq:spd}).  

\begin{figure}[tb]
  \begin{center}
    \includegraphics[width=0.8\linewidth , clip] {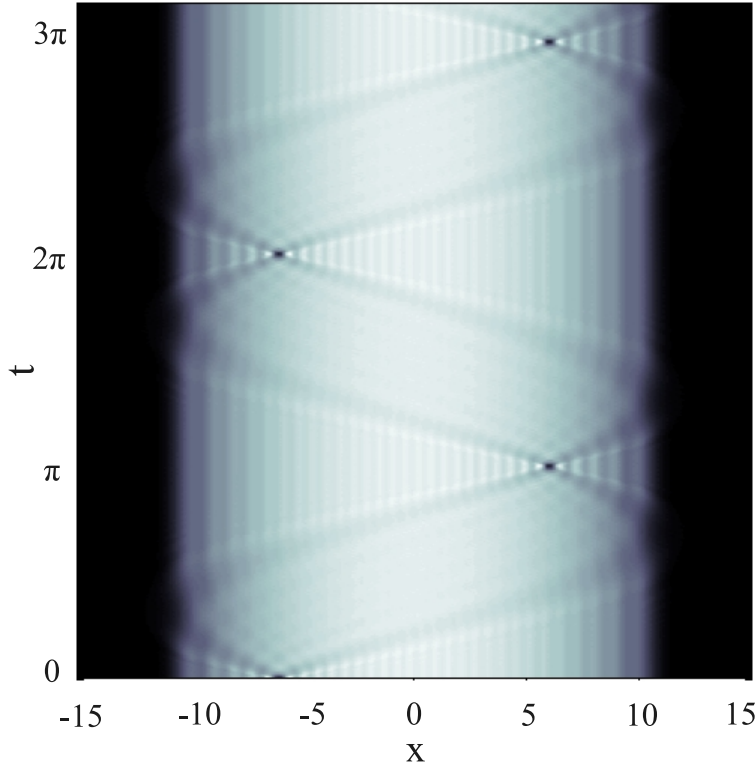}
  \end{center}
  \caption{ The time dependent single particle density of a
    harmonically trapped Tonks-Girardeau gas following the sudden
    removal of a $\delta$ barrier of strength $\kappa=10$, located at
    position $d=-6$.}
  \label{fig:spd_time}
\end{figure}

\begin{figure}[tb]
  \begin{center}
    \includegraphics[width=0.8\linewidth , clip] {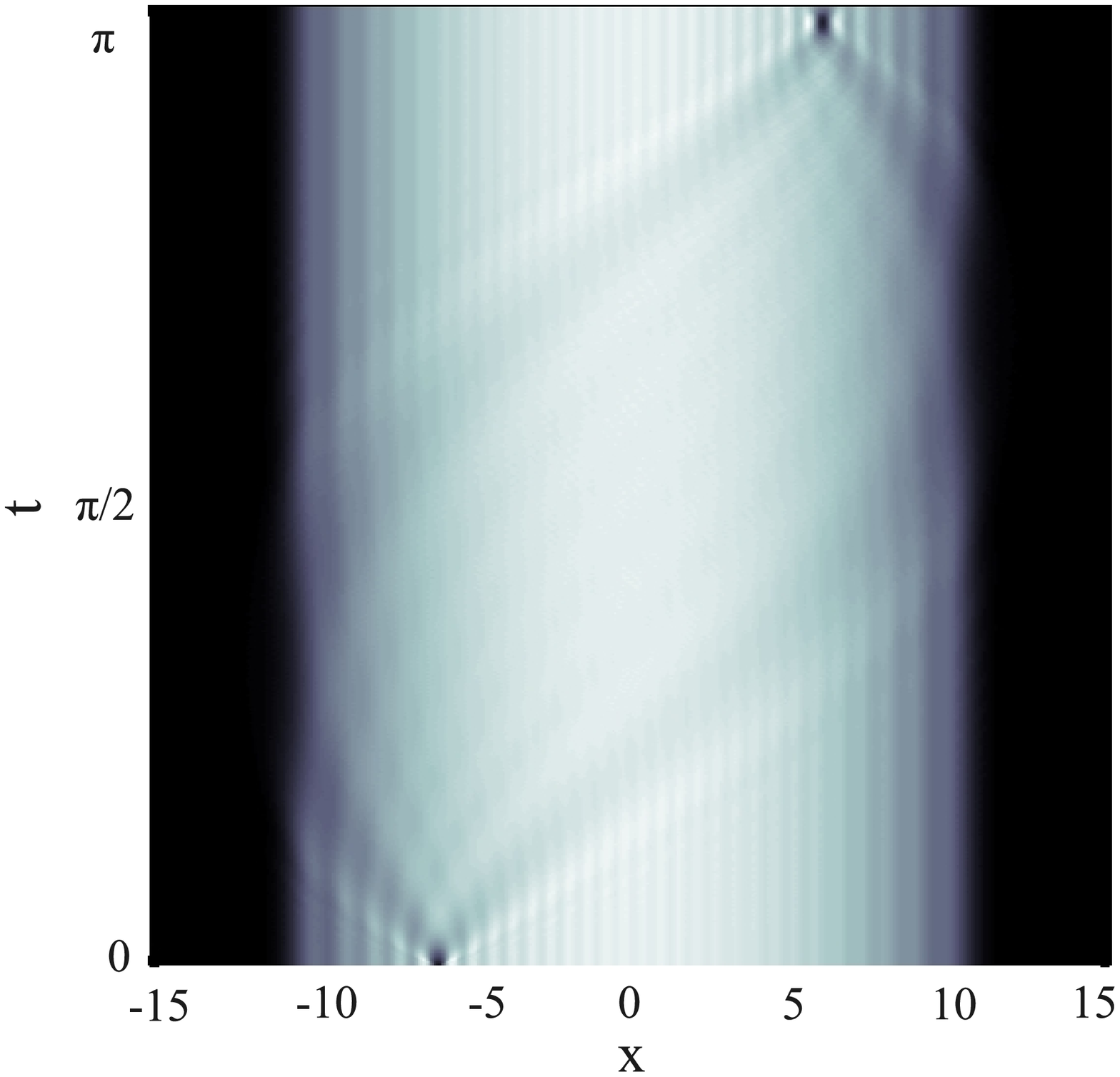}
  \end{center}
  \caption{ Detailed view of the time evolution of the single particle
    density for half a trap period.}
  \label{fig:spd_time_2}
\end{figure}

In Fig.~\ref{fig:spd_time} we show $\rho(x,t)$ following the sudden
removal of an impurity of strength $\kappa=10$ located at $d=-6$. One
can see that the density dip formed by the impurity vanishes almost
instantaneously, however a mirror image of the dip appears after half
a trap period, $t=\pi/\omega$, and then again dissapears followed by a
complete revival after one trap period, $t=2\pi/\omega$. This effect
is analogous to the Talbot effect from classical optics where periodic
refocusing of a diffraction grating is expected to occur in the near
field of a transmitted wave. In this situation the $\delta$-function
represents the most trivial form of a diffraction grating. A natural
question to ask is why is such an effect occurring in a strongly
correlated \textit{many} body system as the Talbot effect is a
coherent \textit{single} particle effect? The simple answer can be
found by recalling that the system can be mapped onto free fermions,
for which the single particle density is simply the sum of the squares
of the single particle eigenfunctions, with each one undergoing its
own coherent unitary evolution. In this picture, the occurrences can
be explained by noting that all $N$ eigenfunctions superimpose in
phase again after every trap period and one finds that the density
profile at odd multiples of $\pi/\omega$ is a mirror image of the
initial density profile at $t=0$. It is also worth noting that in
between this revivals the density shows an interesting fine
structure. This is shown in a close up of the density for the time
period $0\le t\le\pi$ in Fig.~\ref{fig:spd_time_2}. When the impurity
is removed, the matterwave readjusts to the profile of a harmonically
trapped gas by a relaxation of the Friedel oscillations. At $t=\pi/2$
there is complete relaxation of the oscillations, this is followed by
a complete revival at $t=\pi$. This is precisely the fine structure we
see in between revivals of the density dip in Fig.~\ref{fig:spd_time}.

\subsection{Energy profile}

\begin{figure}[tb]
  \begin{center}
    \includegraphics[width=0.8\linewidth , clip] {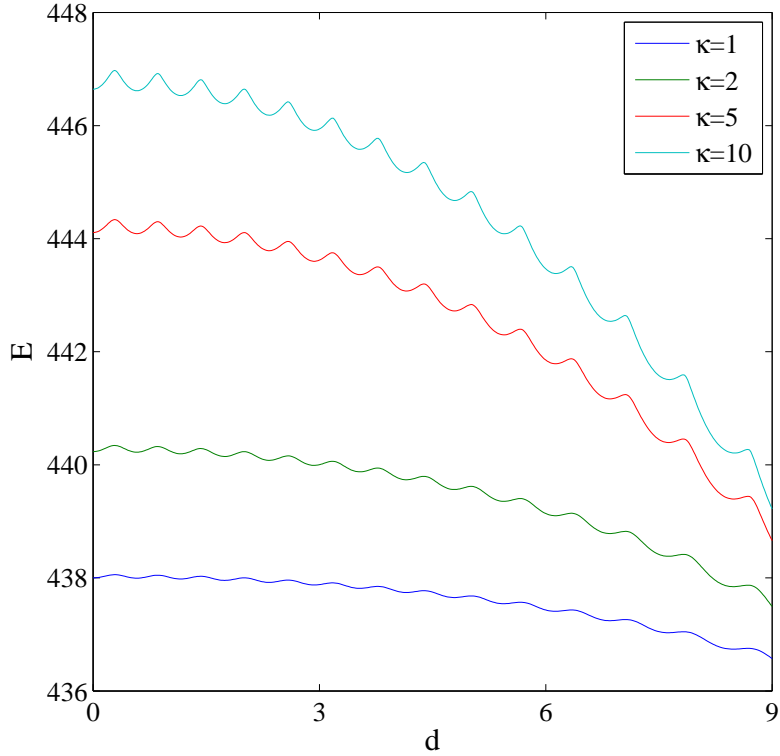}
  \end{center}
  \caption{The energy of a harmonically trapped Tonks-Girardeau gas of
    N=30 particles as a function of the position of the
    $\delta$-perturbation.}
  \label{fig:e}
\end{figure}

It is also interesting to consider the energy of the system as a
function of the position of the impurity. Due to the Fermi-Bose
duality, this can simply be calculated adding the eigenenergies of the
single particle states up to the Fermi energy
\begin{equation}
  \label{eq:ef}
  E_{TG}=\sum_{n=0}^{N-1}\epsilon_{n}\;.
\end{equation}
The energy profile for a gas of thirty particles for different
$\delta$-perturbation strengths as a function of the position of the
perturbation is shown in Fig.~\ref{fig:e}. One can see a series of
exactly fifteen lobes (the plot is symmetric for $d<0$.), which become
more pronounced as the strength of the perturbation is increased, but whose position is independent of this increase. The
position of these local maxima correspond to the positions in which
the probabilities for the single particle wavefunctions peak,
highlighting the crystal structure of the ground state. The increase
in energy due to the impurity is largest when it is located at the
centre of the cloud. In fact, one can view the perturbation as a probe
which, when dragged adiabatically through a Tonks gas, makes it
possible to gain information on the crystal structure of the state
through an energy measurement. In the next section we will look at
non-local properties of the ground state properties, which can be
calculated from the reduced single particle density matrix.

\subsection{Reduced single particle density matrix}
The calculation of the reduced density matrix,
\begin{eqnarray}
\rho^{1}(x,x',t)=N\int^{\infty}_{-\infty}&\Psi_B^*(x,x_2,\dots,x_N,t)\times
                \nonumber\\
                &\Psi_B(x',x_2,\dots,x_N,t)dx_{2}\dots dx_{N}\;,
\label{eq:rspdm_t} 
\end{eqnarray}
and related observables for an ultracold gas is, in general, a
difficult feat. The dimension of the integral in
eq.~(\ref{eq:rspdm_t}) increases with particle number and this is very
demanding on computer memory resources. For a Tonks gas in a harmonic
potential, studies have therefore mainly used numerical methods such
as Monte-Carlo integration to calculate the RSPDM \cite{Triscari:01},
but some analytic work has also been done in this direction
\cite{Lapeyre:02}. Recently, an exceptionally efficient algorithm for
calculating the RSPDM of a Tonks-Girardeau gas in an arbitrary
external potential has been presented by Pezer and Buljan
\cite{Pezer:07}. This algorithm allows for a numerically exact
calculation of the RSPDM for a large number of particles with limited
memory resources and at a rapid computational speed. The algorithm
works for both time dependent and time independent potentials. The
essential idea is that $\rho^1(x,x',t)$ can be expressed in terms of
the dynamically evolving single particle energy eigenbasis,
$\psi_i(x,t)$, as
\begin{equation}
  \rho^1(x,x',t)=\sum^N_{i,j}\psi_i^*(x,t)A_{ij}(x,x',t)\psi_j(x',t)\;.
  \label{eq:rspdm_algorithim} 
\end{equation}
The $N\times N$ matrix, $\textbf{A}(x,x',t)$, is given by
\begin{equation}
  \textbf{A}(x,x',t)=(\textbf{P}^{-1})^{T} \det  \textbf{P}\;,
  \label{eq:A} 
\end{equation}
where the entries of the matrix $\textbf{P}$ are computed as
\begin{equation}
  P_{ij}(x,x',t)=\delta_{ij}-2\int^{x'}_{x}d\xi\psi^*_i(\xi,t)\psi_{j}(\xi,t)\;,
  \label{eq:P0} 
\end{equation}
and where $\delta_{ij}$ is the Kronecker delta. Given a pair of points
$(x,x')$ and the single particle basis functions $\psi_{i}(x,t)$ one
can calculate the RSPDM of a Tonks gas by merely calculating an
$N\times N$ matrix $P$, its inverse and its determinant, which is a
significant saving on computational resources.

\begin{figure}[tb]
  \begin{center}
    \includegraphics[width=\linewidth , clip] {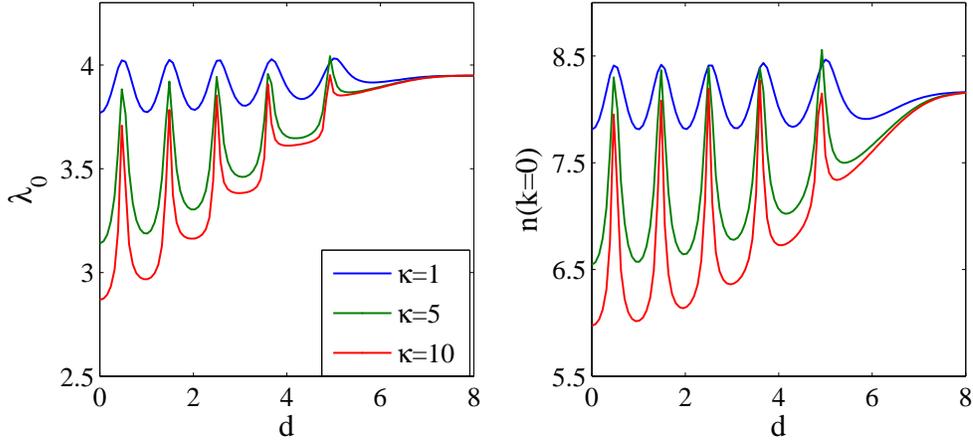}
  \end{center}
  \caption{The largest eigenvalue of the RSPDM $\lambda_{0}$ and the
    peak of the momentum distribution $n(k=0)$ as a function of
    eccentricity $d$ for perturbation strengths $\kappa=1,5, 10$ for a
    Tonks-Girardeau gas of $N=10$ particles.}
  \label{fig:lam_n_10}
\end{figure}
\begin{figure}[tb]
  \begin{center}
    \includegraphics[width=\linewidth , clip] {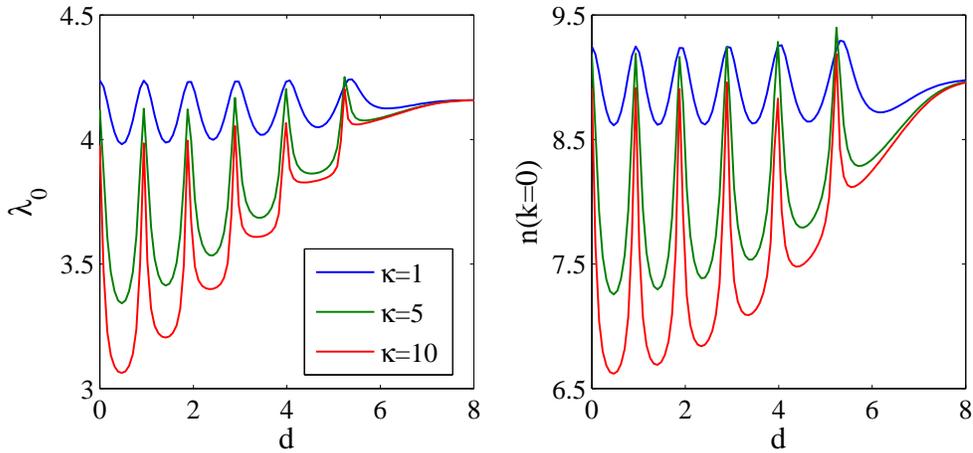}
  \end{center}
  \caption{The largest eigenvalue of the RSPDM $\lambda_{0}$ and the peak of the momentum distribution $n(k=0)$ as a function of eccentricity $d$ for perturbation strengths $\kappa=1,5, 10$ for a Tonks-Girardeau gas of $N=11$ particles. }
  \label{fig:lam_n_11}
\end{figure}

The RSPDM is a hermitian matrix and one can therefore write its
spectral decomposition as
\begin{equation}
  \rho^1(x,x',t)=\sum_{i=0}\lambda_i(t)\phi^*_i(x,t)\phi_i(x',t)\;.
  \label{eq:rspdmdecomp}
\end{equation}
The eigenvectors, $\phi_i$, are often referred to as natural orbitals
in many-body physics and they provide an alternative basis of
orthonormal single particle states for the many-body system. The
eigenvalues, $\lambda_{i}$, are known as the occupation numbers
obeying the normalisation condition $\sum_i \lambda_i=N$. The
expansion in eq.~(\ref{eq:rspdmdecomp}) is extremely useful for
understanding the ground state properties of cold atomic gases, as the
natural orbitals are defined not only for an ideal gas but also for
interacting, thermal and non-uniform gases. In order to obtain the
eigenvectors and eigenvalues of the reduced density matrix, one must
solve the integral value equation,
\begin{equation}
  \int dx'\rho^1(x,x',t)\phi_{i}(x',t)=\lambda_{i}(t)\phi_i(x,t)\;.
\end{equation}
The fraction of particles that are in the lowest lying orbital
$\phi_0(x,t)$ is related to the largest eigenvalue $\lambda_0$ of the
RSPDM by $f=\frac{\lambda_0}{N}$. Therefore, in analogy to the
macroscopic occupation of a single eigenstate in a Bose-Einstein
condensate, this orbital is sometimes referred to as the {\sl BEC}
state and the quantity $\lambda_0$ can act as a measure of the
coherence in the system.

\begin{figure}[h]
  \begin{center}
   \includegraphics[width=\linewidth , clip]{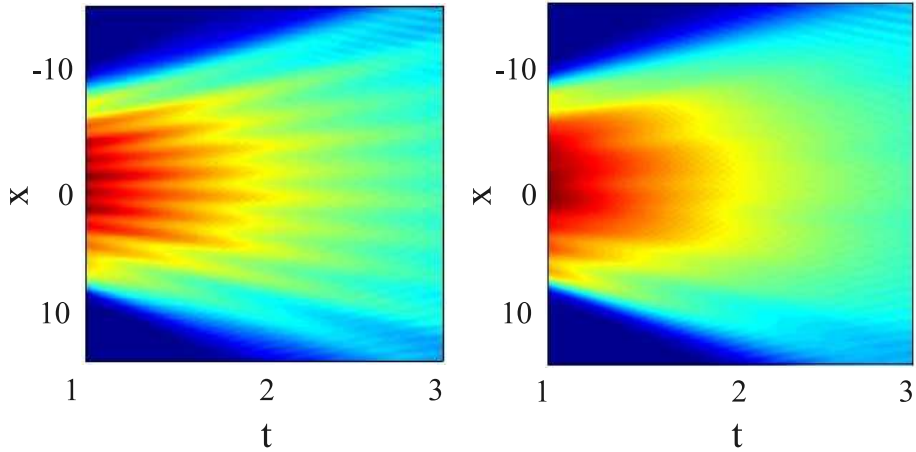}
\end{center}
\caption{Single particle density of the free evolution of the gas
  after realease from the trapping potential. Two cases are shown, (a)
  the delta barrier is positioned at $d=-0.5$ corresponding to a
  maximum coherence and (b) where the delta barrier is positioned at
  $d=-1$ corresponding to minimum coherence. Interference fringes
  indicate greater spatial coherence which confirms predictions of
  Fig.\ref{fig:lam_n_10}.}
\label{fig:interf}
\end{figure}

Another important measure of coherence can be derived from the
momentum distribution of the gas which is defined as
\begin{eqnarray}
n(k,t)&=(2\pi\hbar)^{-1}\int\int\rho^1(x,x',t) e^{\frac{ik(x-x')}{\hbar}}dx\; dx'\\
      &=\sum_i\lambda_i(t)|\mu_i|^2\;,
      \label{eq:momdist}
\end{eqnarray}
where $\mu_{i}=(2\pi\hbar)^{-\frac{1}{2}}\int dx\;
e^{\frac{ikx}{\hbar}}\phi_{i}(x,t)$ are the Fourier transforms of the
natural orbitals. For a homogenous, non-interacting Bose gas at zero
temperature the momentum function is a $\delta$-function, reflecting
the macroscopic occupation of the lowest natural orbital, whereas in
the strongly interacting Tonks-Girardeau gas in equillibrium, the
momentum distribution is comprised of a non-trivial distribution of
quasi-momenta. The amplitude of the peak of the momentum distribution
at $k=0$ can therefore also be used to measure the spatial coherence
present in the system. It is well known that this quantity does not
follow a trivial behaviour in a disturbed, strongly interacting gas
and that it in particular can show a dependence on having an even or
an odd number of particles \cite{Goold1:08,yin:08,lelas:09,yin:10}. In
Fig.~\ref{fig:lam_n_10} we show both quantities defined above,
$\lambda_0$ and $n(k=0)$, as a function of distance of the
perturbation from the trap centre for gas of $N=10$ particles. As
expected, both quantities exhibit similar features dominated by an
oscillatory structure which becomes more pronounced as the strength of
the $\delta$-perturbation increases. As in the interpretation of the
energy oscillations, the maxima correspond to positions where the
single particle eigenstates have large probabilities, i.e. where the
position of the $\delta$-potential corresponds to a lattice point of
the underlying crystal structure of the Tonks gas. For comparison in
Fig.~\ref{fig:lam_n_11} we show the same quantities for the $N=11$
case. In this case we see a maximum of coherence at $d=0$, as symmetry
reasons require a single particle to sit in the centre of the trap for
odd particle numbers \cite{Goold1:08}. This dependence of the
coherence on the position of the disturbance and can be experimentally
observed by measuring the visibility in an interference experiment.
We show in Fig.~\ref{fig:interf} the single particle density for a gas
of $N=10$ particles for which the disturbance is either located at a
maximum of coherence (left panel), or at a minimum of the coherence
(right panel). The difference in the visibility of the interference
fringes is clearly observable.

\section{ Conclusions}
\label{sect:conclusions}
In conclusion we have introduced a model to describe the harmonically
trapped Tonks-Girardeau gas in the presence of a $\delta$-perturbation
of arbitary strength and eccentricity. We have clearly outlined the
solution of the fundamental single particle problem and shown how it
can be used in combination with known techniques to describe the
groundstate properties of the gas. We have calculated the energy
profile of the gas in the presence of the impurity and found an
undulating profile as the perturbation is displaced through the gas,
which highlights the crystal structure of the Tonks groundstate. In
addition, we have calculated the momentum distribution and largest
eigenvalue of the RSPDM as a function of both the eccentricity and
strength of the perturbation. Again, we have found that these
properties reflect the highly localised nature of the particles in the
Tonks gas. Furthermore as a novel application of our model we have
investigated the time density dynamics after a sudden removal of the
perturbation. We find that the gas exhibits the classical Talbot
effect with the image of the impurity reappearing at multiples of the
inverse trap frequency. Given the recent experiments on out-of
equillibrium dynamics of a Tonks gas \cite{Kinoshita:06,Kohl:09}, it
is an interesting extension of our work to study the dynamics of a
moving impurity. This is currently work in progress.

\section{Acknowledgements}
  This work was supported by the National Research Foundation and
  Ministry of Education, Singapore and Science Foundation Ireland
  under grant No.~05/IN/I852. JG would like to thank Elica Kyoseva for
  interesting discussions and MK would like to thank Agnieszka
  Witkowska for invaluable inspiration. TF acknowledges support from IRCSET under the Embark Initiative No.RS/2009/1082. ZI acknowledges financial support from Polish Government Research Grant for years 2007-2010. TB would like to thank Mauro
  Ferreira for helping to stimulate this work.


\section*{Bibliography}
\begin{thebibliography}{}

  % Many-body physics with ultracold gases
\bibitem{Bloch:08} I.~Bloch, J.~Dalibard and W.~Zwerger,
  Rev.~Mod.~Phys.~{\bf 80}, 885 (2008).
 
  % Observation of a Feshbach Resonance in Cold Atom Scattering
\bibitem{Courteille:98} Ph.~Courteille, R.S.~Freeland, D.J.~Heinzen,
  F.A.~van Abeelen and B.J.~Verhaar, Phys.~Rev.~Lett.~\textbf{81}, 69
  (1998).
  
  % Quantum phase transition from a superfluid to a Mott insulator in
  % a gas of ultracold atoms
\bibitem{Greiner:02} M.~Greiner, O.~Mandel, T.~Esslinger,
  T.W.~H\"ansch and I.~Bloch, Nature \textbf{415}, 39 (2002).
 
  % Cold Bosonic Atoms in Optical Lattices
\bibitem{Jaksch:98} D.~Jaksch,C.~Bruder, J.I.~Cirac, C.~Gardiner and
  P.~Zoller, Phys.~Rev.~Lett.~\textbf{81}, 3108, (1998).
 
  % Berezinskii–Kosterlitz–Thouless crossover in a trapped atomic gas
\bibitem{Hadzibabic:06} Z.~Hadzibabic, P.~Kr\"uger, M.~Cheneau,
  B.~Battelier and J.~Dalibard, Nature \textbf{441}, 1118 (2006).
   
  % Relationship between Systems of Impenetrable Bosons and Fermions
  % in One Dimension
\bibitem{Girardeau:60} M.~Girardeau, J.~Math.~Phys.~\textbf{1}, 516
  (1960).

  % Tonks–Girardeau gas of ultracold atoms in an optical lattice
\bibitem{Parades:04} B.~Paredes, A.~Widera, V.~Murg, O.~Mandel,
  S.~F\"olling, I.~Cirac, G.V.~Shlyapnikov, T.W.~H\"ansch and
  I.~Bloch, Nature \textbf{429}, 277 (2004).

  % Observation of a One-Dimensional Tonks-Girardeau Gas
\bibitem{Weiss:04} T.~Kinoshita, T.~Wenger and D.S.~Weiss, Science
  \textbf{305}, 1125 (2004).
  
  % A quantum Newton's cradle
\bibitem{Kinoshita:06} T.~Kinoshita, T.~Wenger and D.S.~Weiss, Nature
  \textbf{440}, 900 (2006).

  % Quantum Transport through a Tonks-Girardeau Gas
\bibitem{Kohl:09} S.~Palzer, C.~Zipkes, C.~Sias and M.~K\"ohl,
  Phys.~Rev.~Lett.~{\bf 103}, 150601 (2009).
  
  % Low-density, one-dimensional quantum gases in a split trap
\bibitem{Busch:03} Th.~Busch and G.~Huyet,
  J.~Phys.~B:~At.~Mol.~Opt.~\textbf{36}, 2553 (2003).

  % Tonks-Girardeau gas in an optical lattice: Effects of a local
  % potential
\bibitem{Fu:06} H.~Fu and A.G.~Rojo, Phys.~Rev.~A \textbf{74}, 013620
  (2006).
  
  % Ground State Properties of a Tonks-Girardeau Gas in a Split Trap
\bibitem{Goold1:08} J.~Goold and Th.~Busch, Phys.~Rev.~A {\bf 77},
  063601 (2008).
  
  % Low-density, one dimensional quantum gases in the presence of a
  % localised attractive potential
\bibitem{Goold2:08} J.~Goold, D.~O'Donoghue and Th.~Busch,
  J.~Phys.~B:~At.~Mol.~Opt.~\textbf{41}, 215301 (2008).
  
  % Motion of an impurity particle in an ultracold
  % quasi-one-dimensional gas of hard-core bosons
\bibitem{Girardeau:09} M.~Girardeau and A.~Minguizzi, Phys.~Rev.~A
  {\bf 79}, 033610 (2009).
  
  % Ion induced density bubble in a strongly correlated one
  % dimensional gas
\bibitem{Goold:10} J.~Goold, H.~Doerk, Z.~Idziaszek, T.~Calarco and
  Th.~Busch, Phys.~Rev.A {\bf 81}, 041601 (2010).
  
  % Exciting Collective Oscillations in a Trapped 1D Gas
\bibitem{Moritz:03} H.~Moritz, T.~St\"oferle, M.~K\"ohl, and
  T.~Esslinger, Phys.~Rev.~Lett.~\textbf{91}, 250402 (2003).

 
  % Atomic Scattering in the Presence of an External Confinement and a
  % Gas of Impenetrable Bosons
\bibitem{Olshanii:98} M.~Olshanii, Phys.~Rev.~Lett.~\textbf{81}, 938
  (1998).
  
  % Beautiful Models
\bibitem{Sutherland} B.~Sutherland, {\sl Beautiful Models}, (World
  Scientific Publishing Company, 2004).  
 
  % Boson pairs in a one-dimensional split trap
\bibitem{Murphy:07} D.S.~Murphy, J.F.~McCann, J.~Goold and Th.~Busch,
  Phys.~Rev.~A \textbf{76}, 053616 (2007).

  % Two Atoms in seperate harmonic traps
\bibitem{Krych:09} M.~Krych and Z.~Idziaszek, Phys.~Rev.~A {\bf 80},
  022710 (2009).

  % Handbook of Mathematical Functions
\bibitem{Abramowitz:72} M.~Abramowitz and I.~Stegun, {\sl eds.}, {\sl
    Handbook of Mathematical Functions} (Dover, 1972).
  
  % Breakdown of Time-Dependent Mean-Field Theory for a
  % One-Dimensional Condensate of Impenetrable Bosons
\bibitem{Girardeau:00} M.~Girardeau and E.M.~Wright,
  Phys.~Rev.~Lett.~\textbf{84}, 5239 (2000).

  % Metallic alloys
\bibitem{Friedel:58} J.~Friedel, Nouvo Cimento \textbf{7}, 287 (1958).

  % Quantum Mechanics and Path Integrals
\bibitem{Feynman} R.P.~Feynman and A.R.~Hibbs, {\sl Quantum Mechanics
    and Path Integrals}, (McGraw-Hill, New York, 1965).
  
  % Ground-state properties of a one-dimensional system of hard-core
  % bosons in a harmonic trap
\bibitem{Triscari:01} M.D.~Girardeau, E.M.~Wright, and J.M.~Triscari,
  Phys.~Rev.~A \textbf{63}, 033601 (2001).

  % Momentum distribution for a one-dimensional trapped gas of
  % hard-core bosons
\bibitem{Lapeyre:02} G.J.~Lapeyre, Jr., M.D.~Girardeau, and
  E.M.~Wright, Phys.~Rev.~A \textbf{66}, 023606 (2002).
  
  % Momentum Distribution Dynamics of a Tonks-Girardeau Gas: Bragg
  % Reflections of a Quantum Many-Body Wave Packet
\bibitem{Pezer:07} R.~Pezer and H.~Buljan,
  Phys.~Rev.~Lett.~\textbf{98}, 240403 (2007).
  
  % Ground-state properties of a few-boson system in a one-dimensional
  % hard-wall split potential
\bibitem{yin:08} X.~Yin, Y.~Hao, S.~Chen, and Z.~Zhang, Phys.~Rev.~A
  \textbf{78}, 013604 (2008).
  
  % Ground-state properties of a one-dimensional strongly interacting
  % Bose-Fermi mixture in a double-well potential
\bibitem{lelas:09} K.~Lelas, D.~Juki\'c and H.~Buljan, Phys.~Rev.~A
  \textbf{80}, 053617 (2009).
  
  % Hard-core Bose-Fermi mixture in one-dimensional split traps
\bibitem{yin:10} X.~L\"u, X.~Yin and Y.~Zhang, Phys.~Rev.~A
  \textbf{81}, 043607 (2010).
  
  
\end{thebibliography}
\end{document}